Central and Peripheral Vision for Scene Recognition: A Neurocomputational Modeling Exploration

Panqu Wang

Department of Electrical and Computer Engineering, University of California, San Diego

La Jolla, CA, USA

pawang@ucsd.edu

Garrison W. Cottrell

Department of Computer Science and Engineering, University of California, San Diego

La Jolla, CA, USA

gary@ucsd.edu

# Abstract

What are the roles of central and peripheral vision in human scene recognition? Larson and Loschky (2009) showed that peripheral vision contributes more than central vision in obtaining maximum scene recognition accuracy. However, central vision is more efficient for scene recognition than peripheral, based on the amount of visual area needed for accurate recognition. In this study, we model and explain the results of Larson and Loschky (2009) using a neurocomputational modeling approach. We show that the advantage of peripheral vision in scene recognition, as well as the efficiency advantage for central vision, can be replicated using state-of-the-art deep neural network models. In addition, we propose and provide support for the hypothesis that the peripheral advantage comes from the inherent usefulness of peripheral features. This result is consistent with data presented by Thibaut et al. (2014), who showed that patients with central vision loss can still categorize natural scenes efficiently. Furthermore, by using a deep mixture-of-experts model ("The Deep Model", or TDM) that receives central and peripheral visual information on separate channels simultaneously, we show that the peripheral advantage emerges naturally in the learning process: When trained to categorize scenes, the model weights the peripheral pathway more than the central pathway. As we have seen in our previous modeling work, learning creates a transform that spreads different scene categories into different regions in representational space. Finally, we visualize the features for the two pathways, and find that different preferences for scene categories emerge for the two pathways during the training process.

# Introduction

Viewing a real-world scene occupies the entire human visual field, but visual resolution across the visual field varies dramatically. Foveal vision, for example, extends to about 1° of eccentricity from the center of the visual field (Polyak, 1941), within which the highest density of cones (Curcio, Sloan, Kalina, & Hendrickson, 1990; Wandell, 1995) and highest spatial resolution (Hirsch & Curcio, 1989; Loschky et al., 2005) are found. Next to the foveal region is the parafoveal region, which has slightly lower spatial resolution and extends to about 4-5° eccentricity (Rayner et al., 1981; Coletta & Williams, 1987), where high density of rods is found. Beyond the parafovea is commonly referred as peripheral vision (Holmes, Cohen, Haith, & Morrison, 1977; van Diepen, Wampers, & d'Ydewalle, 1998), where the retina has the highest proportion of rods and the lowest spatial resolution.

Central (foveal and parafoveal) vision and peripheral vision serve different roles in processing visual stimuli. Due to the high density and small receptive field of foveal photoreceptors, central vision encodes more fine-detailed and higher resolution information compared to peripheral vision, which encodes coarser and lower spatial frequency information (Sasaki et al., 2001; Musel et al., 2013). This suggests that recognition processes requiring high spatial frequency usually favor central vision more than peripheral vision, as in object and face recognition. Behavioral studies have shown that object recognition performance is the best within 1° to 2° of eccentricity of the fixation point, and performance drops rapidly as eccentricity increases (Henderson & Hollingworth, 1999; Nelson & Loftus, 1980). For face recognition, studies have shown that face identification performance is also severely impaired in peripheral vision (Harry, Davis, & Kim, 2012), which is presumably caused by the reduced spatial acuity in the periphery (Melmoth, Kukkonen, Mäkelä, & Rovamo, 2000) or crowding (Martelli, Majaj, & Pelli, 2005). Studies of scene recognition, however, suggest that low spatial frequencies and global layout play a key role in recognizing scene gist (McCotter, Gosselin, Sowden, & Schyns, 2005; Loschky et al., 2007; Sanocki, 2003). As a result, it is natural to argue that peripheral vision plays a more important role in scene recognition.

In addition to the behavioral studies, brain imaging studies have shown that orderly central and peripheral vision representations can be found not only in low-level retinotopic visual areas (V1 to V4), but also in high-level visual areas in ventral temporal cortex, when perception and recognition for faces or scenes is being performed (Levy et al., 2001; Malach, Levy, & Hasson, 2002; Hasson, Harel, Levy, & Malach, 2003; Grill-Spector & Malach, 2004, Arcaro, McMains, Singer, & Kastner, 2009). More specifically, these studies argue that cortical topography, particularly eccentricity mapping, is the underlying principle of the organization of the higher order visual areas: objects whose recognition relies more on fine detail, such as words and faces, are associated more with central representations; recognition that relies more on global shape and large-scale integration, as in the case of buildings and scenes, is associated with peripheral representations. This hypothesis is supported by fMRI evidence that shows that the brain regions that respond more strongly to faces (Fusiform Face Area, or FFA, Kanwisher, McDermott, & Chun, 1997) and words (Visual Word Form Area, or VWFA, McCandliss, Cohen, & Dehaene, 2003) than other categories are associated with central representations, whereas the regions that are more activated by buildings and scenes (Parahippocampal Place Area, or PPA, Epstein, Harris, Stanley, & Kanwisher, 1999) sit in the eccentricity band enervated by the peripheral visual field (Grill-Spector & Weiner, 2014). Nasr et al. (2011) showed that scene-selective areas of human visual cortex (e.g., PPA, retrosplenial cortex (RSC), and Occipital Place Area (OPA)) tend to have a peripheral vision bias, with RSC and PPA immediately adjacent to peripheral representations of V1 and V2, and V2 to V4, respectively. Baldassano, Fei-Fei, & Beck (2016) identified a bias in functional connectivity to peripheral V1 throughout scene-sensitive regions, and demonstrated that functional correlations during natural viewing reflect eccentricity biases in high-level visual areas. More recent studies suggest that the central-biased face recognition pathway and peripheral-biased scene recognition pathway are functionally and anatomically segregated by the mid-fusiform sulcus (MFS) to enable fast and parallel processing of categorization tasks in the ventral temporal cortex (Grill-Spector & Weiner, 2014; Weiner et al., 2014; Gomez et al., 2015; Lorenz et al., 2015).

Given the above hypothesis that central and peripheral representations are distinct anatomically, and that peripheral vision is associated with scene recognition, it is natural to ask the question, "What are the relative contributions of central versus peripheral vision in scene recognition?" Larson and Loschky (2009) performed a behavioral study using a "Window" and "Scotoma" design (See Figure 1) to address this question. In the Window condition, human subjects viewed scene images through a circular window centered on the fovea. The image within the window is unaltered, while the image outside the window is absent. The Scotoma condition is just the reverse of the Window condition, with a central circular area blocked and the outside region unaltered. The Window and Scotoma paradigm has been applied to various studies in scene perception (Loschky & McConkie, 2002; Loschky et al., 2005; Larson, Freeman, Ringer, & Loschky, 2014), as it provides a way to understand the importance of information in a particular area: if the missing information is important, the normal information processing flow will be disrupted and performance will be impacted; if the missing information is not needed, there should not be any difference in performance.

In the study of Larson and Loschky (2009), 100 human subjects were recruited to perform a scene categorization task for 10 categories (5 Natural: Beach, Desert, Forest, Mountain, and River; 5 Man-made: Farm, Home, Market, Pool, and Street), using four different sets of eccentricity radii (1°: foveal, 5°: central, 10.8°: equal viewable area, and 13.6°: large window), under Window and Scotoma conditions. For each of the 320 self-paced trials in the experiment, subjects were first presented a flashed scene, then were asked to choose "Yes" or "No" based on whether the stimulus matched the post-cue word. The results of Larson & Loschky (2009) are summarized in Figure 3 (a) and (d). First, they found that central (foveal+parafoveal) vision is not necessary for recognizing scenes, although it contains high resolution details that are very important for face and object recognition. Conversely, peripheral vision, despite its low resolution, is important for scene recognition to achieve maximum accuracy. In addition, they also found that central vision is more efficient than peripheral vision on a per-pixel basis; when the visual area shown is equalized between the two conditions, less central area is needed to achieve equal accuracy. The crossover point, where central vision starts to outperform peripheral, is less than 10.8°. In an additional experiment, they found a critical radius of 7.4° where the Window and Scotoma conditions produce equal performance. They found this empirical critical radius value is significantly larger than the value predicted by V1 cortical magnification equations (Florack, 2000; Van Essen, Newsome, & Maunsell, 1984), and suggested that the utility of central vision for scene recognition is less than would be predicted by V1 cortical magnification.

In this work, we explain the data in Larson and Loschky (2009), using a neurocomputational modeling approach. More specifically, we aim to answer the following questions: Can we use a brain-inspired model to replicate the behavioral data? If so, can the model provide any insights on why peripheral vision contributes more to scene recognition than central vision? Can the model explain how the peripheral advantage emerges from a development perspective? Finally, what are the differences between central and peripheral representations?

We answer these questions using deep convolutional neural network (CNN)-based models. First, we show that our modeling results match the observations of Larson and Loschky (2009). Second, we suggest that the peripheral preference for scene recognition emerges from the inherent usefulness of the peripheral features: A model trained using only peripheral vision outperforms a model trained using only central vision. This hypothesis is supported by the findings of Thibaut et al. (2014), who showed that people with central vision loss can still efficiently categorize natural scenes. Eberhardt, Zetzsche, & Schil (2016) further showed that peripheral features are especially more useful for scene localization and scene categorization tasks, but not for object recognition, where foveal features are more important. Third, we used a deep mixture-of-experts model ("The Deep Model", or TDM) to demonstrate how a pathway that receives peripheral visual information gradually gains an advantage over a pathway that receives only central visual information: When the two are in competition, the peripheral pathway learns a transform that differentiates the scene categories in its representational space faster than the central pathway. This suggests that there is a natural developmental reason for the peripheral pathway to become the scene recognition system. Finally, we use a simple method to visualize the learned features in our model, and find that the central and peripheral pathway have different preferences over the scene categories.

## Methods

**Image Dataset**

We obtained images from the ten categories of stimuli (but not the same images) that were used in Larson & Loschky's behavioral study from the Places205 Database (Zhou et al., 2014), which contains 205 different scene categories and over 2.5 million images. The ten classes we used have a total of 129,210 training images that range from 7278 (for the pool category) to 15,000 (for 6 out of 10 categories) images per category, and 1000 test images (100 per category). All input images were preprocessed using the retina model described in the next section. As ten categories is a relatively small number, and may cause overfitting issues when training a deep CNN from scratch, we used models pre-trained on the full Places205 Database[1] and performed fine-tuning (or transfer learning) on the ten categories based on these models. One can think of the pretrained models that are already able to perform scene recognition tasks, as modeling a mature scene recognition system in the brain. The fine-tuning process, however, is just additional training on a new but similar task, similar to the subjects' practice trials in the behavioral study. In addition, it is required to adapt the network to our log-polar-transformed images.

---

[1] Downloaded from http://places.csail.mit.edu/downloadCNN.html

**Image Preprocessing**

In our experiments, we used two types of input images: foveated images and log-polar transformed images. Given the raw images have the same spatial resolution across the whole image, the foveated representation of the images mimics the human retina by varying the spatial resolution across input images based on eccentricity. To create foveated images, the Space Variant Imaging System[2] was used. To mimic human vision, the parameter that specifies the eccentricity where spatial resolution drops to half of the center of fovea is set to 2.3° throughout the experiments (Geisler & Perry, 1998). To further account for the fact that mapping between retina and the cortex in human visual system is a log-polar transformation that creates cortical magnification of central representations (Schwartz, 1977; Rojer & Schwartz, 1990, Wilkinson, Anderson, Bradley, & Thibos, 2016), we apply log-polar transforms on the foveated images. Log-polar transformation has been applied in computational models, such as modeling the retina (Bolduc & Levine, 1998), performing active object recognition (Kanan, 2013), and modeling the determination of the focus of expansion in optical flow at different retinal eccentricities (Chessa, Maiello, Bex, & Solari, 2016). We use the well-established OpenCV method[3] to generate log-polar transformed images, where the scale parameters are $M = width/log(radius_{max})$, and $radius_{max} = \sqrt{2} \cdot width/2$. An example scene image, the preprocessed foveated version, and the log-polar transformed version is shown in Figure 1.

The images were further processed to match the Window and Scotoma paradigms, as in the experiments of Larson and Loschky (2009). All input images in our experiments have a dimension of $256 \times 256$ pixels; we assume this corresponds to $27° \times 27°$ of visual angle, consistent with Larson and Loschky (2009). As described in the previous section, four Window and Scotoma set of radii are used in the behavioral study: $1°$ for the presence or absence of foveal vision; $5°$ for the presence or absence of central vision; $10.8°$ equates the viewable area (in pixels) inside and outside the Window/Scotoma; $13.6°$ presents more area within the Window/Scotoma than outside. In our modeling study, we added five additional radii to make the predictions of the model more precise, namely $3°$, $7°$, $9°$, $12°$, and $16°$. The example Window and Scotoma images are shown in Figure 1. Note for $5°$, the Scotoma condition contains much larger area (number of pixels) than the Window condition using foveated image (Window:Scotoma=1:8.4); however, the Scotoma condition has a much smaller area than the Window condition using log-polar transformed images (Window:Scotoma=1:0.34), due to the effect of cortical magnification.

---
[2] http://svi.cps.utexas.edu/software.shtml
[3] goo.gl/3i2WOS

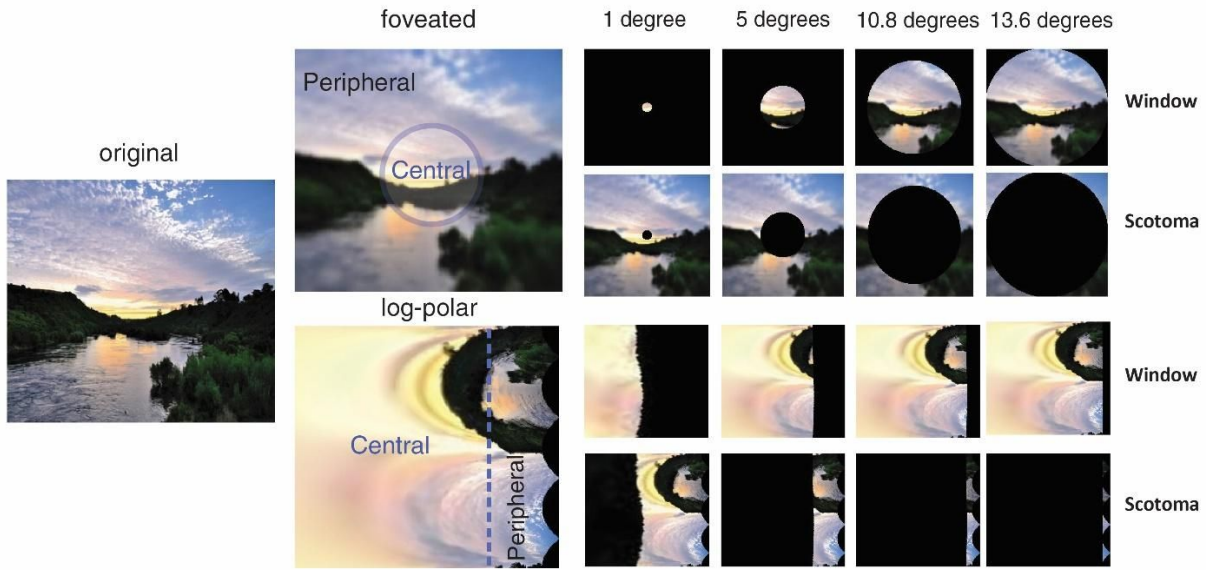

**Figure 1.** Example of an image used in our experiment. First column: original image. Second column: foveated image and log-polar transformed image. Third column to last column: the foveated and log-polar transformed images processed under Window and Scotoma conditions with different radii in degrees of visual angle.

**Deep Convolutional Neural Networks (CNNs)**

All of the models we used in our experiments are based on deep CNNs. Deep CNNs are neural networks that have (many) more layers than the traditional multi-layer perceptron, with a computational hierarchy that repeatedly stacks the following operations: 1) a set of learned, 2-dimensional convolutions, performed on either the input stimulus or the feature responses from the previous layer; 2) spatial pooling, which is applied to a small local region of the feature maps obtained from the convolution operation, which reduces the dimensionality of the feature map and gains some degree of translational invariance; 3) nonlinear activation functions applied to the pooled responses. The nonlinearity gives the network its power, as otherwise it would simply be a linear system (modulo the max pooling). As layers deepen, the receptive fields of the learned filters generally become larger as they receive input from the pooled responses of the previous layer. The learned filters at early layers become low-level features (edges, corners, contours), while later layers become high-level object-related representations (object parts or entire objects) (Zeiler & Fergus, 2014). On top of these stacked computations are usually some fully connected layers that combine features from across the entire image and are more abstract and task-dependent. Finally, the output layer represents the target categories, typically as a softmax layer with a cross-entropy objective function so that the output is the probability of the category, given the input (Bishop, 1995).

We used the deep CNN framework in our experiments for two main reasons. First, deep CNNs are the current state of the art in computer vision, as they have achieved the best performance on numerous large-scale computer vision tasks, such as object recognition (Krizhevsky, Sutskever, & Hinton, 2012; He, Zhang, Ren, & Sun, 2015), object detection (Ren, He, Girshick, & Sun, 2015), video classification (Karpathy et al., 2014), and scene recognition (Zhou, Lapedriza, Xiao, Torralba, & Oliva, 2014; Shen, Lin, & Huang, 2015). Leveraging the representation learned using millions of parameters from millions of training examples, deep CNNs are becoming competitive with or better than human performance on various tasks, such as traffic sign classification (Cireşan, Meier, Masci, & Schmidhuber, 2012) and face recognition (Taigman, Yang, Ranzato, & Wolf, 2014). As a result, deep CNN based models should achieve reasonable performance in our experiments. Smaller networks or other algorithms are not competent for our tasks, given their relatively weaker generalization power compared with deep CNNs. Second, deep CNNs have been shown to be excellent models of the primate visual cortex, as they are able to predict a variety of neural data in monkey and human IT (Cadieu et al., 2014; Yamins et al., 2014; Agrawal, Stansbury, Malik, & Gallant, 2014; Wang, Malave, & Cipollini, 2015; Yamins & DiCarlo, 2016). For example, Güçlü & van Gerven (2015) have demonstrated that deep CNNs achieve the state-of-the-art decoding performance from the blood oxygenation level-dependent signal (BOLD) response in the ventral stream, and the learned features quantitatively match the observations in Zeiler and Fergus (2014). As a result, it is a natural choice to use CNN-based approaches in modeling a behavioral study related to human vision.

The exact CNN models we used in our experiments vary according to the experimental setting and the task. In the experiments modeling the behavioral data of Larson and Loschky (2009), in order to investigate whether different network structures, especially depth variation, give different results in the modeling task, we applied three popular feed-forward single pathway architectures in our experiment, which are shown as follows:

1. AlexNet (Krizhevsky et al., 2012), which contains 5 convolutional layers and 3 fully connected layers. The network has approximately 60 million trainable parameters, and achieves $81.10\%$ top-5 accuracy (the correct category is in the top five responses of the network) on the Places205 validation set.
2. VGG-16 Net (Simonyan & Zisserman, 2014), which contains 13 convolutional layers and 3 fully connected layers. The network has approximately 138 million trainable parameters, and achieves $85.41\%$ top-5 accuracy on the Places205 validation set.
3. GoogLeNet (Szegedy et al., 2015), which contains 21 convolutional layers and 1 fully connected layer. The network has approximately 6.8 million trainable parameters, and achieves $87.70\%$ top-5 accuracy on the Places205 validation set.

In our experiments, a two-pathway (one central and one peripheral) CNN model (namely "The Deep Model", or TDM) using a mixture-of-experts architecture (Jacobs, Jordan, & Barto, 1991) is used, which will be explained in the next section.

**The Deep Model (TDM)**

The idea of TDM is based on its ancestor, "The Model" (TM; Dailey & Cottrell, 1999; Cottrell & Hsiao, 2011) (Figure 2(b)). In TM, each input stimulus is processed through two biologically-plausible preprocessing layers: 2-D Gabor filtering, simulating the response of V1 complex cells, and principal component analysis (PCA), which reduces the dimensionality of the gabor filter responses and models the information extraction process beyond the primary visual cortex. After these steps, the feature vector is fed into a neural network with two side-by-side hidden layers that adaptively learn the features for a given task. For example, if the task is face (subordinate classification) and object recognition (basic level categorization), we can consider the two hidden layers as corresponding to the FFA (Fusiform Face Area) and the LOC (Lateral Occipital Complex). A softmax gating layer is imposed to modulate the learned weights from the hidden layer to the output layer based on the relative contributions of the two modules: if one module is better at processing a given pattern, the gating layer will direct more information (error feedback) through the node corresponding to that module by increasing the value of that gating node. TM has been used to model and explain many cognitive processes, such as the development of hemispheric lateralization in face processing (Dailey & Cottrell, 1999; Wang & Cottrell, 2013), and why the FFA is recruited for non-face categories of expertise (Tong, Joyce, & Cottrell, 2008; Wang, Gauthier, & Cottrell, 2016).

TDM is an extension of TM to deep CNNs. As deep CNNs are usually trained from end to end (pixels to labels), neither Gabor filtering nor the PCA preprocessing step is needed in TDM. Rather, we can simply build two deep CNNs as the two modules in TM.

Compared with previous deep CNN-based modeling studies (Yamins & DiCarlo, 2016, Güçlü & van Gerven, 2015), we are the first to adopt a mixture-of-expert architecture in neurocomputational modeling work, to the best of our knowledge. In our experiment, one component represents the pathway that is enervated by central vision, and the other represents the pathway that is enervated by peripheral vision. The two components process their input in parallel until the last fully connected layer, which connects to the output layer. The gating layer connects the output weights of the two layers to the final layer, which hypothetically contains discriminative information between central and peripheral vision. The TDM model is illustrated in Figure 2(c). The TDM in this instance simulates the fact that central and peripheral representations are segregated by the mid-fusiform sulcus in ventro-temporal cortex (Grill-Spector & Weiner, 2014; Weiner et al., 2014; Gomez et al., 2015), but hypothesizes that these two sources of information are integrated in an anterior area to make the final categorization decision.

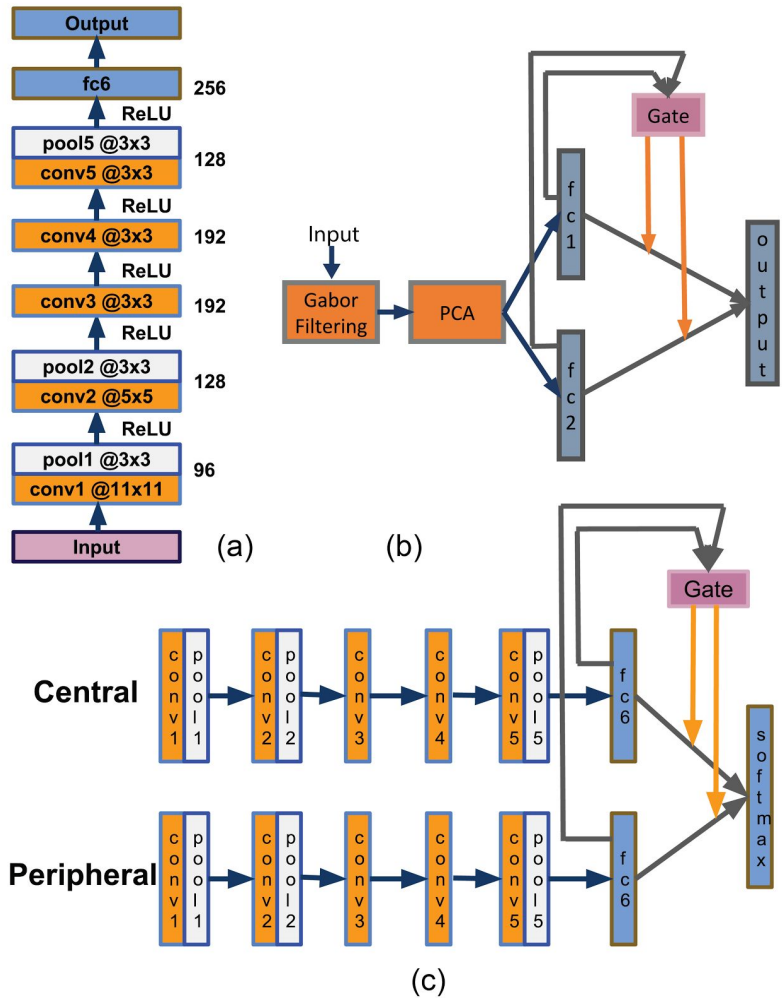

**Figure 2.** The network architecture of TM and TDM. (a) shows the structure of one the two pathways in TDM; this network is used in Experiment 3.1. The network has 7 layers, including 5 convolutional layers with filter size $M{\times}M$ and $N$ (the number to the right of each layer) feature maps for each layer, and 2 fully connected (fc6 and output) layers. (b) shows the architecture of TM. The input is preprocessed by Gabor filter banks and PCA before feeding into a two-layer neural network, and the output layer is modulated by the gating layer (Gate). (c) shows the two-pathway TDM (used in Experiment 3.2) that models central and peripheral visual information processing. The two side-by-side pathways have identical structure, and converge at the output layer, with the weights between *fc6* and the output layer modulated by the gating layer (Gate).

# Results

In this section, we first present the experiments modeling the behavioral data in Larson and Loschky (2009). We then use our CNN model to explain the model results.

**Experiment 1: Modeling Larson and Loschky (2009)**

Larson and Loschky's (2009) measured the relative contribution of central versus peripheral vision in scene recognition; the results are summarized in Figure 3(a). The Scotoma condition outperforms the Window condition for the $1°$ (foveal vision) and $5°$ (central vision) visual angle settings. This suggests that losing central vision (the Scotoma condition) does not severely impair scene recognition performance, but losing peripheral vision (Window condition) does. As a result, peripheral vision is more important than central vision for attaining maximal scene recognition performance. However, central vision is more efficient than peripheral vision on a per-pixel basis, because performance in the Window condition is better than in the Scotoma condition when the presented areas are equal ($10.8°$). This is best illustrated when the data is plotted as a function of percentage of viewable area, as shown in Figure 3(d).

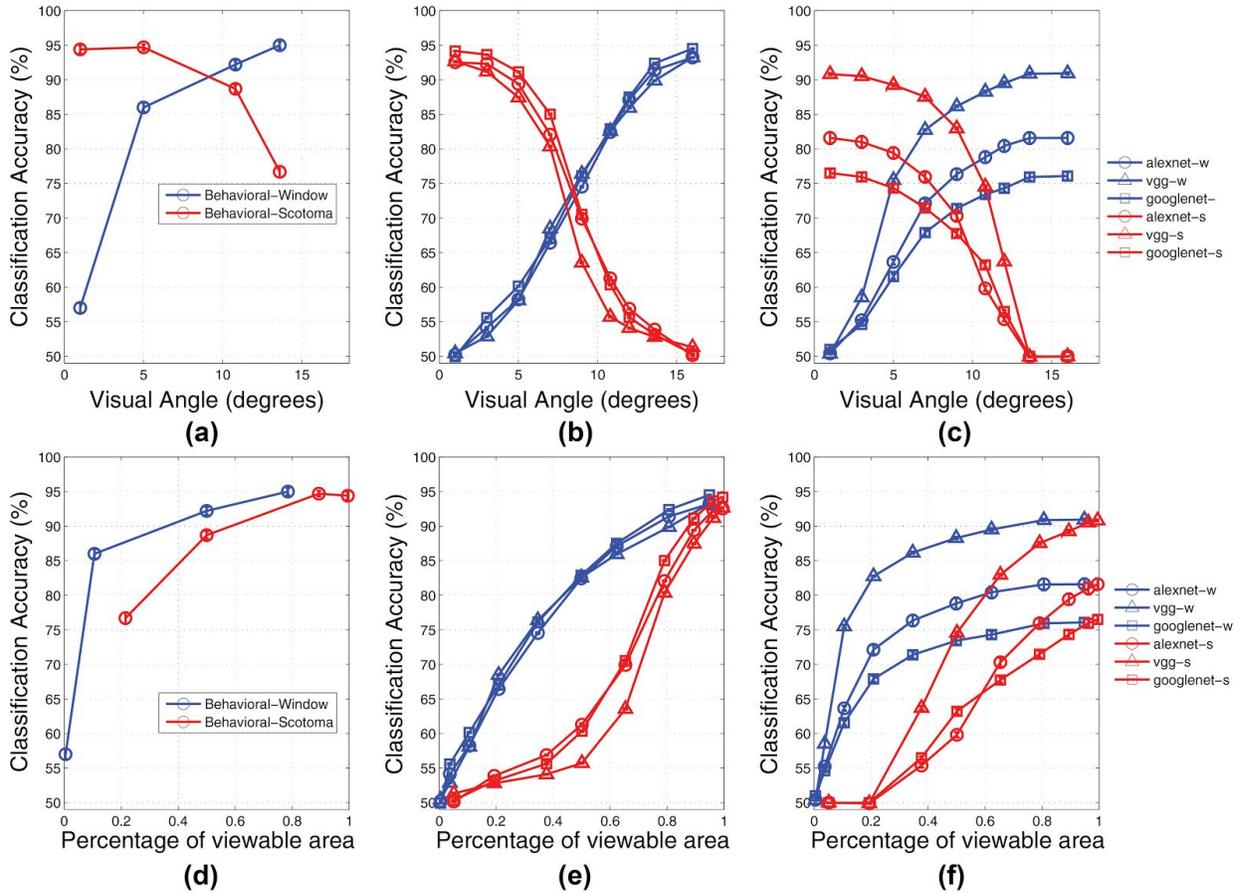

**Figure 3.** Summary of the behavioral study results in Larson and Loschky (2009) and our modeling results. Top row: scene recognition accuracy as a function of viewing condition (Windows (w) and Scotomas (s)). (a) result of behavioral study; (b) model results using foveated images; (c) model results using log-polar transformed images. Bottom row: results for scene recognition accuracy as a function of viewable area for Window and Scotoma conditions. (d) result of behavioral study; (e) model results using foveated images; (f) model results using log-polar transformed images. Each data point is obtained by averaging over 20 "subject" networks. Some standard error bars are invisible in the graph.

In our experiment, we have two parallel settings: one using foveated images only, and one that additionally applies the log-polar transform to the foveated images to account for cortical magnification. The hypothesis is that cortical magnification will weigh central vision more than peripheral vision, and thus may eliminate the peripheral advantage. In the fine-tuning process for all models, we initialized the weights in all but the last fully connected layer to their pretrained values, and initialized the weights of the last layer randomly with zero mean and unit variance. To be consistent with the behavioral study in which the subject is performing a verification task (e.g., "is this a pool scene?", as opposed to "which category is this?"), we used 10 logistic units rather than a softmax, so each output is independent of the others. For each output unit, we trained it using half of the images from the target category (positive examples) and half of the images randomly selected from all 9 other categories (negative examples). We set the learning rate of the last layer to be $0.001$ as it needs to learn faster, and all other layers to be $1e^{-4}$ as they need only minor adjustments from their pre-trained state. Weight decay was set to $5e^{-4}$, and momentum was set to be 0.9. All models were fine-tuned using mini-batch gradient descent with batch size of 256 (AlexNet and GoogLeNet) or 32 (VGG-16, due to memory constraints), running on an NVIDIA Titan Black 6GB GPU, using the Caffe deep learning framework (Jia et al., 2014). We trained all networks for a maximum of 24,000 iterations to ensure convergence. This is probably overkill for the experiment using foveated images, as the data is similar to a subset of the training data. For the experiment using log-polar transformed images, however, more training is necessary, as the appearance of the image is completely different from the original training data (Figure 1). The test set for each category contained 200 images, 100 from the target category and 100 randomly chosen (but not used in training) from the other nine categories. All test images were preprocessed to meet with each of the Window and Scotoma conditions. The test accuracy is defined as mean classification accuracy across all categories. The results are shown in Figure 3. Note that since this is a yes/no decision, chance is 50%.

From Figure 3, we can clearly see that our modeling results show the same trends as the behavioral data, for both log polar and foveated images. The characteristics shared by both transforms are: First, as the radius of visual angle (x axis) increases, the classification accuracy increases or decreases monotonically for the Window and Scotoma conditions. Second, the Scotoma condition yields better performance than the Window condition for central vision (less than $5°$), consistent with the results of Larson and Loschky (2009). Third, we also replicated the result that central vision is more efficient than peripheral vision, as the Window condition outperforms Scotoma condition in the radius of equal viewable areas ($10.8°$). When we plot the performance of central and peripheral vision as a function of viewable areas (Figure 3(e)(f)), we can clearly see that central vision achieves better performance than peripheral vision for all conditions and all models we tested. This finding also matches the behavioral results (Figure 3(d)), which show that the central vision is superior to peripheral as the number of viewable pixels increase. These results suggest that our models are quite plausible.

In addition, our results are consistent with a recent behavioral study that showed people with age related macular degeneration (AMD, or central vision loss) can still categorize scenes efficiently (Thibaut et al., 2014). Patients with AMD were still able to categorize scenes using low-frequency based peripheral vision, although normally sighted controls performed better than patients with AMD. Our results are consistent with these findings.

However, the shapes of the curves in the log-polar condition are in stark contrast to those in the foveated condition. They are much closer to the behavioral data with the log polar transform, especially for the VGG-16 network. The log-polar transform also leads to more variance in performance between the different networks, which behave quite similarly under the foveating image transformation. While the 16-layer network is in the middle, depth-wise, one not so obvious difference is that, due to differences in architectural details, VGG-16 actually has many more parameters than the other two networks: 138M compared to AlexNet's 60M and GoogLeNet's 6.8M. The log-polar transform dramatically distorts the images compared to what these networks were initially trained on, whereas simply foveating them does not. With many more parameters, VGG net has more flexibility to adapt to the log polar transform than the other two networks, resulting in dramatically better performance.

This change in format would also explain why the classification accuracy of the log-transformed images is slightly lower than the behavioral study. If we started from initial weights and retrained these models from scratch, we presumably would have better results in accuracy. In addition, humans have a great deal of prior experience with occlusion, while the networks have not. Models with more realistic experience, both with log-polar images from the start, and occlusion, may be needed to fully account for the results in the behavioral study.

However, even without these changes, we see that VGG-16 with log-transformed images displays performance curves that are much more in line with the behavioral data, compared to networks using foveation alone. First, in the Window condition (blue curves), the model using log-polar images shows a much more rapid increase in classification accuracy as the visual angle increases from 1 to 5 degrees, similar to the human subjects, which is clearly due to cortical magnification. Second, somewhat counterintuitively, for the Scotoma condition, the decrease of classification of accuracy as the degree of the scotoma increases is much slower compared to merely foveated images, which again, fits the behavioral data better. The explanation for this lies in another fact about the log-polar transform: The logarithmic representation means that as the degree of the scotoma increases, the amount of input to the peripheral network drops more slowly, resulting in less disruption. Looking at Figure 1, the reduction in visual area in the foveated version between 5° and 10° is relatively greater than the reduction in visual area in the log-polar version. This phenomenon again demonstrates how incorporating realistic anatomical constraints into computational models provides better explanations of (and fits to) the data.

**Photographer Bias:** One may argue that our result may be influenced by photographer bias (Schumann et al., 2008; Tatler, 2007), in that our training and testing images are taken in a stereotypical way, and more information concerning scene category is located in the center of the image. Photographer bias has a potential risk of contaminating computational modeling results (Tseng et al., 2009; Zhang, et al., 2008). Consequently, we examined whether our modeling result is affected by photographer bias. The procedure is similar to Experiment 1, except that we now located the center of the fovea at 4 different locations: ($h$/4,$w$/4), ($h$/4,$3w$/4), ($3h$/4,$w$/4), and ($3h$/4,$3w$/4), where $h$ and $w$ are the height and width of the image, respectively. We also placed the center of the Window or the Scotoma at the location of each new fovea, and performed log-polar transformation based on the new center. This "four centers" configuration to test the photographer bias is similar to that in Velisavljević & Elder (2008), except that we did not crop the image at each location. Same as Experiment 1, we tested 9 Window/Scotoma pairs across different visual eccentricity settings on test images. Since the result obtained from the four different locations hardly differs, we averaged the data from these locations. Our results are shown in Figure 4.

From Figure 4, we can see that the general trend we observed in Figure 3(b) still holds: peripheral vision is more useful than central vision to obtain the best scene recognition accuracy. All networks show the same behavior, with the VGG-16 network still outperforming the others. This result demonstrates that the peripheral advantage seen in our model is not affected by photographer bias.

However, while the peripheral advantage remains, there exists nuances between the performance of the two models. First, for the Window conditions, the classification accuracy is generally lower if we move the fovea away from the center to the four off-center locations (for example at 5°, the accuracy for VGG is over 75% for the center, but only around 65% for the quadrants). This result suggests that there exist more important features centered in the photos than the off-center locations used for scene recognition. In addition, for the Scotoma conditions, the classification accuracy drops more slowly in the off-center conditions than in the center condition. This is because the "periphery" in the Scotoma conditions for off-center locations actually include the center of the image, where there are useful features that boost the performance. Overall, we show that photographer bias may play a role in our experiments, but the peripheral advantage is not affected by it.

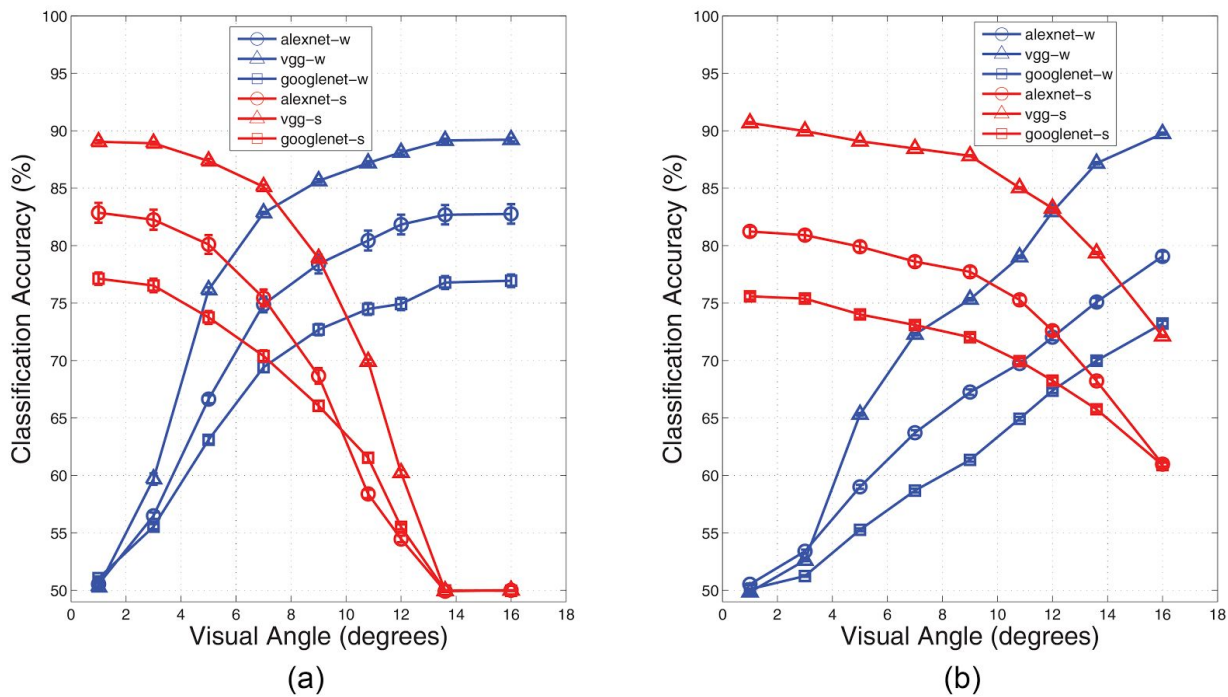

(a)                                (b)

**Figure 4.** Results of testing whether photographer bias exists in our model. (a) A copy of Figure 3(c), where center is located at ($h/2,w/2$) for Window (w) and Scotoma (S) conditions. (b) The averaged result of four different centers: upper left ($h/4,w/4$), upper right ($h/4,3w/4$), lower left ($3h/4,w/4$), and lower right ($3h/4,3w/4$). All networks behave similarly.

**Experiment 2: Critical Radius**

Critical radius is the radius that produces equal scene recognition performance between Window and Scotoma scene images. In Larson and Loschky (2009), they measured the critical radius by testing the recognition accuracy at three candidate radii ($6.0°$, $7.6°$, and $9.2°$), and calculating the crossing point for the two linear equations going through $6.0°$ and $7.6°$ for the Window and Scotoma condition, respectively. 18 human subjects were recruited, and they ran another experiment using the same procedures as their Experiment 1. They found the critical radius is $7.48°$, which is far larger than the predicted critical radii from cortical magnification functions, based on the assumption that equal V1 activation would produce equal performance (Florack, 2007; Van Essen, Newsome, & Maunsell, 1984). They hypothesized that this weaker cortical magnification is possibly due to the greater importance of peripheral vision in the higher order visual areas that subserve scene recognition.

We modeled the behavioral study to find the critical radius for scene recognition in our model. To improve the precision of the prediction, we used 10 different radii that range from $6.0°$ to $9.6°$, with an interval of $0.4°$. As in Larson and Loschky (2009), we use foveated circular images, processed by the log-polar transformation (Figure 5(a)). Again, all three deep CNNs were employed in this experiment, and the training and testing procedures are exactly the same as those in Experiment 1. The result is shown in Figure 5(b).

In Figure 5, we see that the predicted critical radius of the deep CNNs is consistent with the one shown in Larson and Loschky (2009) at the group level: the averaged critical radius across is $8.00°$, and is not significantly different from the result shown in the behavioral study, using a two-tailed t-test ($t$=0.2092, $p$=0.8371). At the individual level, we found the critical radii predicted by VGG-16 Net, GoogLeNet, and AlexNet are $8.25°$, $8.05°$, and $7.70°$, respectively.

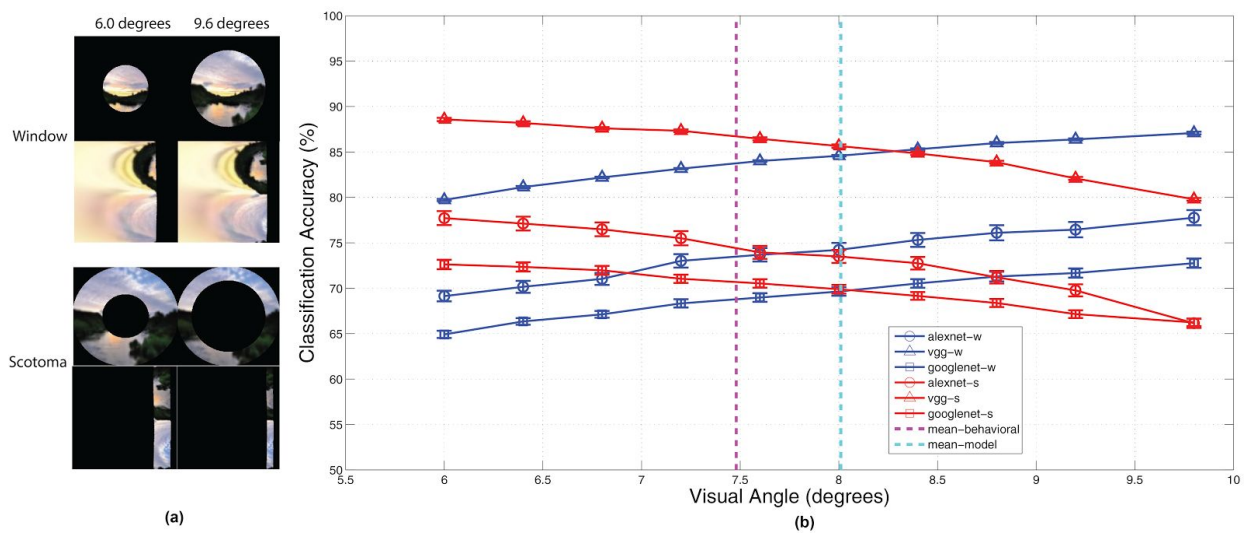

**Figure 5.** Results for modeling the critical radius. (a) example Window and Scotoma

condition images. For each condition we show foveated (first row) and log-polar transformed (second row) version of the same image. Same as Experiment 1, log-polar transformed images are used in this experiment. (b) Scene recognition accuracy between Window and Scotoma radii from $6.0°$ to $9.6°$, are measured for all three models. The predicted critical radius averaged across all three models (the dotted blue vertical line) is $8.00°$. The dotted red vertical line represents the critical radius ($7.48°$) measured in the behavioral study. Each data point is obtained by averaging over 20 "subject" networks. Error bars denote standard errors.

**Experiment 3: Analysis**

We have successfully modeled the main results in Larson and Loschky (2009), using deep CNNs. Our results are consistent with the findings in the behavioral study by demonstrating that peripheral vision is more important than central vision in scene recognition. The next natural question to ask, however, is *why* this is the case, and *how* this might happen in cortex? Here we use our model to provide more insights into this question. In this section, we explain why peripheral vision contributes more to central vision by validating two hypotheses we propose. We also use TDM to illustrate how the peripheral advantage could naturally emerge as a developmental process. We further analyze the features learned by both central and peripheral representations to identify potential differences between the two pathways.

*Why is peripheral vision more important than central vision for scene recognition?*
**Hypothesis 1 (Experiment 3.1):** Peripheral vision contains better features useful for scene recognition than central vision.

The first hypothesis we propose is very simple: peripheral vision simply contains better features for the task than central vision. In our experiment, this implies the visible portion of the image in the Scotoma condition creates better features than that in the Window condition in achieving the maximum scene recognition accuracy. To test this hypothesis, we trained networks with only central or peripheral vision. This differs from the previous experiments in that, training on whole images and then blocking part of them could disrupt processing. In this experiment, we are directly testing whether the information is sufficient for good performance in each case.

We designed an experiment using a deep CNN model to test this hypothesis. Instead of training the model using entire scene images and testing it using Window and Scotoma images as in Experiment 1 and 2, we trained a deep CNN using the Window and Scotoma images directly, and then tested the network using entire scene images. The interpretation of this process is as follows: if peripheral vision contains better feature for scene recognition than central vision, then the network trained under the Window condition will perform more poorly than the network trained under the Scotoma condition at the radius of $5°$, because the loss of peripheral vision incurs more significant loss of accuracy than the loss of central vision. In other words, the network trained using Scotoma images of $5°$ eccentricity will not suffer much loss of accuracy when compared the network trained using entire images, because its loss of central vision is not important, that is, central vision is not that necessary for scene recognition.

We used the same ten scene categories as Larson and Loschky (2009) to run this experiment. Since we now have a much smaller dataset to train the deep CNN (compared to the network pretrained using 205 categories in Experiment 1), the network size must be reduced to alleviate overfitting. We used seven layers in total, with five convolutional layers and two fully connected layers. The total number of trainable parameters is approximately 2.3 million, reduced by almost 96% from AlexNet. The detailed network architecture is shown in Figure 2(a). In this case, the networks' performance was based on classification accuracy, so the output was a 10-way softmax, and chance performance is 10%.

We trained ten networks using different initial random weights at eccentricity radii of $5.0°$. The training set is the same as was used in Experiment 1, except that they were preprocessed to meet with the Window and Scotoma conditions. All networks were trained using mini-batch stochastic gradient descent (SGD) with a batch size of 128 for 30,000 iterations to ensure convergence. The initial learning rate was set to 0.01 and decreased by a factor of 10 every 10000 iterations. Weight decay was set to $5e^{-4}$, and momentum was set to be 0.9. The result is shown in Table 1.

| Data Used | Central | Peripheral | Full |
|---|---|---|---|
| Mean | 0.500 | 0.687 | 0.715 |
| SE | 0.002 | 0.006 | 0.023 |

**Table 1.** Result of scene recognition performance (Mean: averaged classification accuracy across all categories; SE: standard error) on the test set for Experiment 3.1. Central: networks trained using images containing central information only; Peripheral: networks trained using images containing peripheral information only; Full: networks trained using images containing both central and peripheral information. All networks are trained using log-polar transformed foveated images. Chance performance is 0.1.

From Table 1, we can clearly see that for radius $5.0°$, the network trained using peripheral information significantly outperforms the network trained using central information. This result demonstrates that central vision does not provide information as important as peripheral information for scene recognition, thus supporting hypothesis 1. One thing to note, however, is that networks trained using both central and peripheral information still achieve the highest accuracy, indicating that central vision indeed has its own contribution to scene recognition process.

**Hypothesis 2 (Experiment 3.2):** Peripheral vision creates better internal representations for scene recognition than central vision.

Now that we know peripheral vision contains better features than central vision for scene recognition, we further investigate how these features are projected into the representational space of the deep CNNs. To make the comparison of internal representations clear and realistic, we train both networks *simultaneously* using "The Deep Model" (TDM), as normal people receive both central and peripheral vision at the same time. The intuition is that if peripheral vision is more important than central vision, the pathway that represents peripheral vision in TDM should receive more feedback during training, and this will lead to a higher gating value and better internal representations for the scene categories than central vision does.

The architecture of TDM is shown in Figure 2(c). The two pathways have identical network structure and number of parameters, with the network design the same as the previous experiment that tests hypothesis 1. One pathway receives the input that represents central vision using Window images of $5°$, and the other pathway receives the input that represents peripheral vision using Scotoma images of $5°$. The two pathways remain segregated until the last layer that connects to the output, modeling the fact that central and peripheral visual information are processed in parallel and segregated in the ventral temporal cortex (Levy et al., 2001, Malach, Levy, & Hasson, 2002; Grill-Spector & Weiner, 2014). However, information carried in both pathways is integrated when performing categorization in the last layer of the TDM. The location of this integration is hypothesized to be in an anterior region of the ventral temporal cortex, or in the prefrontal cortex. The activations that feed into the gating nodes are from the net input of *fc6* layer, as we assume higher-order semantic information that can be learned from central and peripheral inputs is learned in this layer, and this will help the gating nodes determine the relative contributions of the two pathways. The output of the gating layer is connected with the weights from the last fully connected layer to the output softmax layer. Again, the job of the network here is 10-way classification, so chance is 10%. The total number of free parameters in TDM is approximately 4.62 million.

The training process of TDM is exactly the same as in experiment 3.1. The value of the gating nodes were both initialized to 0.5 to make sure the two networks were initially training equally. After finishing training at iteration 30000, we recorded the value of the gating nodes for both pathways using test images. Note the gating values are influenced by the image - it can *choose* whether to use the central pathway or not on a per-image basis. Hence, the network uses central information only when it is useful. We also included two control conditions in this experiment, in which the network received central information (Window) or peripheral information (Scotoma) in both pathways, respectively. The result is shown in Figure 6.

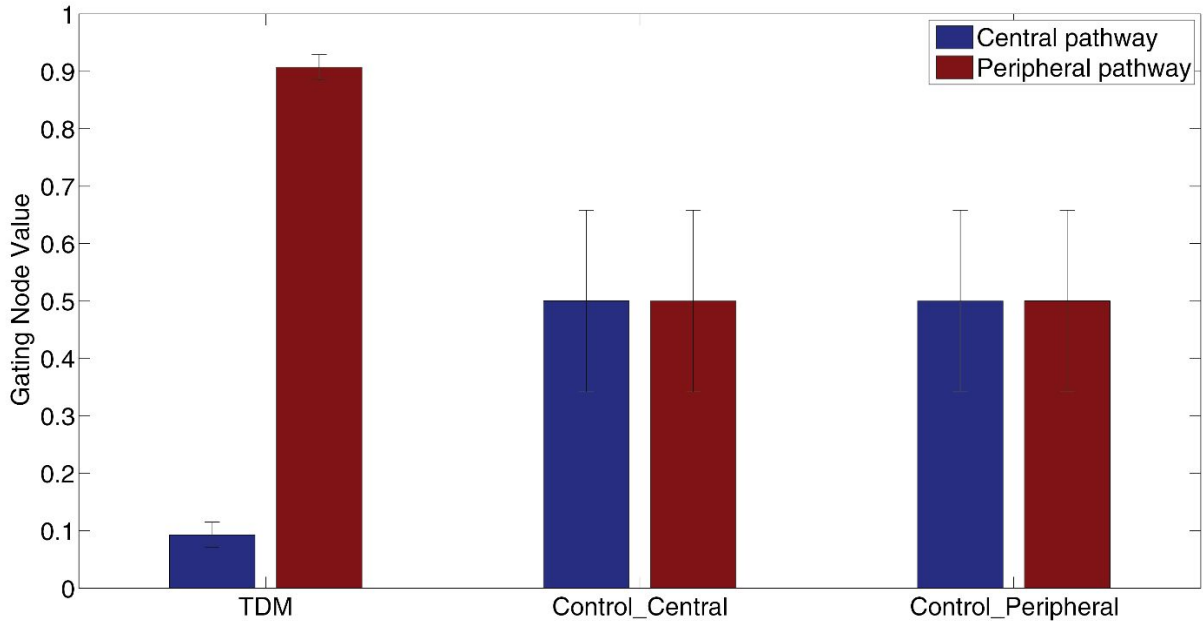

**Figure 6.** Gating node value in TDM for the central pathway (blue bar) and the peripheral pathway (red bar), in this experiment. Control_Central and Control_Peripheral are the two control conditions in that the two networks receive the same central or peripheral information. We can see that the difference between the central and peripheral information causes the difference in gating node value in TDM. All conditions are trained 10 times, and the error bar denotes the standard error.

From Figure 6, we can clearly see that the averaged gating node value for the peripheral pathway ($M = 0.91$, $SD = 0.0697$) is significantly higher than that for central pathway ($M = 0.09$, $SD = 0.0697$) in the experimental condition. There is no significant difference between the gating node values for both control conditions, no matter the training condition. This suggests that the difference between the gating values is caused by the difference between central and peripheral inputs to the model. Our result demonstrates that, given a choice, the network will use the peripheral pathway for scene categorization.

The value of the gating nodes as a function of training iterations is shown in Figure 7. We can see that the advantage of the peripheral representation starts early (around iteration 2000), and becomes stronger as training proceeds. The gating node value for peripheral representation achieved its maximum value around iteration 8000 and then stabilizes. The scene recognition accuracy (the green line in Figure 6) tracks the increasing gating node value for the peripheral pathway. This finding suggests that the peripheral advantage for scene recognition can emerge as a developmental process: although the central and peripheral pathway are equally weighted to begin with, the greater usefulness of the peripheral representation directs the network to learn more information from the peripheral pathway gradually through learning. When the development of both pathways is finished, the weights stabilize, implying the scene recognition system is mature. This process thus can be hypothesized to mimic the development of the scene recognition system in cortex.

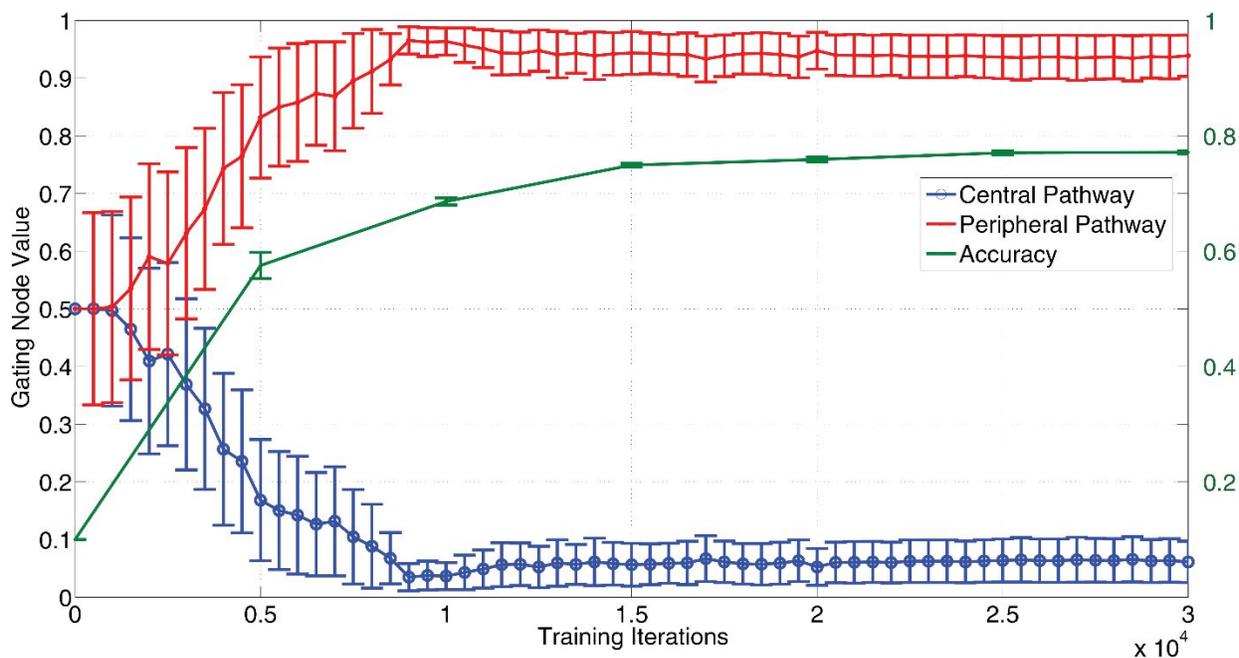

**Figure 7.** Gating node value for central (blue) and peripheral (red) pathway as a function of training iterations. The green line illustrates the scene recognition accuracy of TDM through time. The networks are the same as those used to plot Figure 6.

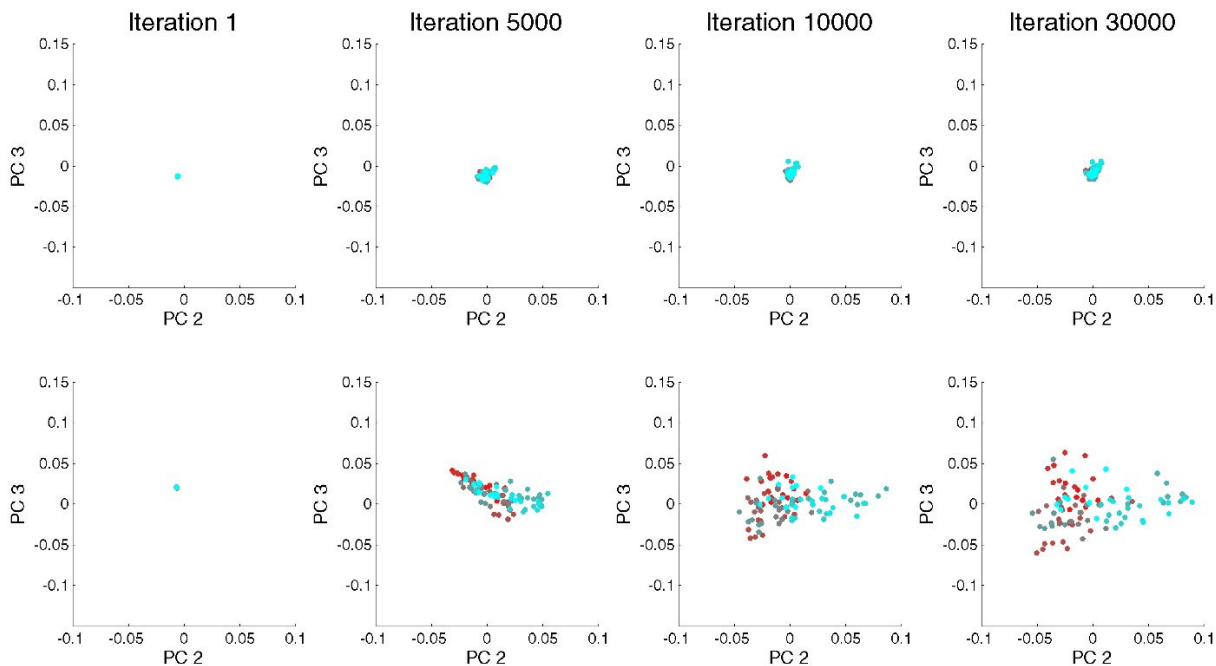

**Figure 8.** Visualization of the development of net input of *fc6* layer over network training of TDM. First row: representations of scene categories in the central pathway. Second row: representations of scene categories in the peripheral pathway. Each column represents the data collected from corresponding training epoch (shown in the title). The colored dots (in 10 different colors, one dot represents one example of a category) represent 10 different object categories used in the experiment.

Besides directly measuring gating node value to probe the relative importance of central and peripheral representations, we can visualize the information contained in the fully connected *fc6* layer in TDM during the training process to gain a better understanding of how the internal representation evolves. Specifically, we collected the net input of the fc6 layer for both pathways across all validation images, at four different time points (iteration 1, 5000, 10000, and 30000). We then performed PCA on the collected data and visualized the projection on the second the third principal components on a 2-D subspace (the first principal component just reflects the magnitude of the activations growing over time). This analysis of the hidden layer activation has proved useful in the past: In work explaining why the FFA is recruited for new object categories of expertise, Tong et al. (2008) showed that fine-level discrimination leads to an expanded representational space that also spreads out new stimuli, while basic level categorization "clumps" objects in representational space, making it difficult to distinguish individual members of a category. In work modeling the effect of experience in face and object recognition, Wang et al. (2016) demonstrated that more experience results in more separation in the hidden unit representational space, and that helps recognize objects of expertise. In this work, we used this technique to analyze the difference between central and peripheral representations generated in TDM. The result is shown in Figure 8.

Distinct patterns between central and peripheral representations can be observed from Figure 8: At iteration 1, central and peripheral representations are the same because learning has not started. At iteration 5000, the peripheral pathway has already learned a transformation that pushes all categories apart from the center, but the central pathway is apparently much worse at doing that: examples from different categories are still squeezed together. At iteration 10000 and 30000, the examples in the peripheral pathway become even more separated in the 2D subspace, but such separation cannot be found in the central pathway. The reason that there is more of a spreading transform by the periphery is that the gating network gives much more error feedback to the peripheral network during training, as a consequence of its initial superior ability -  and this advantage accumulates through training ("rich get richer" effect). Given that our model is a model of scene recognition, we hypothesize that the location of the separation is in the PPA, but more generally, it could be in any area where scene recognition is performed. This is the reason why peripheral vision is more important than central vision for recognizing scenes.

One may argue that the representation developed in the hidden layer of the central pathway is a foregone conclusion, given that it receives very little training feedback from the gating node. Perhaps if it was given more feedback, a better representation would be learned. To overturn this potential objection, we also trained five networks using the same paired architecture, except that we forced the gating weight to remain at 0.5 throughout training for both networks. Remarkably, the network learns to *turn off* the central pathway when it is forced to use central information all of the time - the activations of the ReLU units at the fc6 layer in the central pathway are all 0, for all units, for all five networks trained from scratch. The central pathway units respond (that is, their net input is the least negative) to images that are mostly one color (orange, dark, white, etc.). This is even stronger evidence that, even when given both types of information, gradient descent chooses the peripheral pathway.

Based on the 2D visualization result of the representation in the fc6 layer in the networks with trained gates, we predicted that the features learned in the two pathways must be very different from one another in order to produce this result. We further analyzed the difference between the features learned in both pathways by visualizing them using a technique described in Zhou, Khosla, Lapedriza, Oliva, & Torralba (2015): for *conv1* layer, we visualized the learned weights of all filters by concatenating three RGB channels; for each unit in *pool2* and *fc6* layer, since we cannot visualize the features directly, we selected the top 3 training images that generated the highest activations for that unit. For each of the 3 images, we identified the regions of the image that lead to high unit activation by replicating each image many times with small random occluders at different locations in the image. Specifically, we generated occluders in a dense grid with a patch size of 15 and stride of 4, and this results in about 3600 occluded images per original image. We then fed all occluded images into the same network as the original image, and recorded the difference of the activation value between the occluded images and the original image. If there is a large discrepancy, we know that the corresponding patch is important. We then defined the feature for a given unit as the regions that cause the maximal discrepancies. The example features are shown in Figure 9.

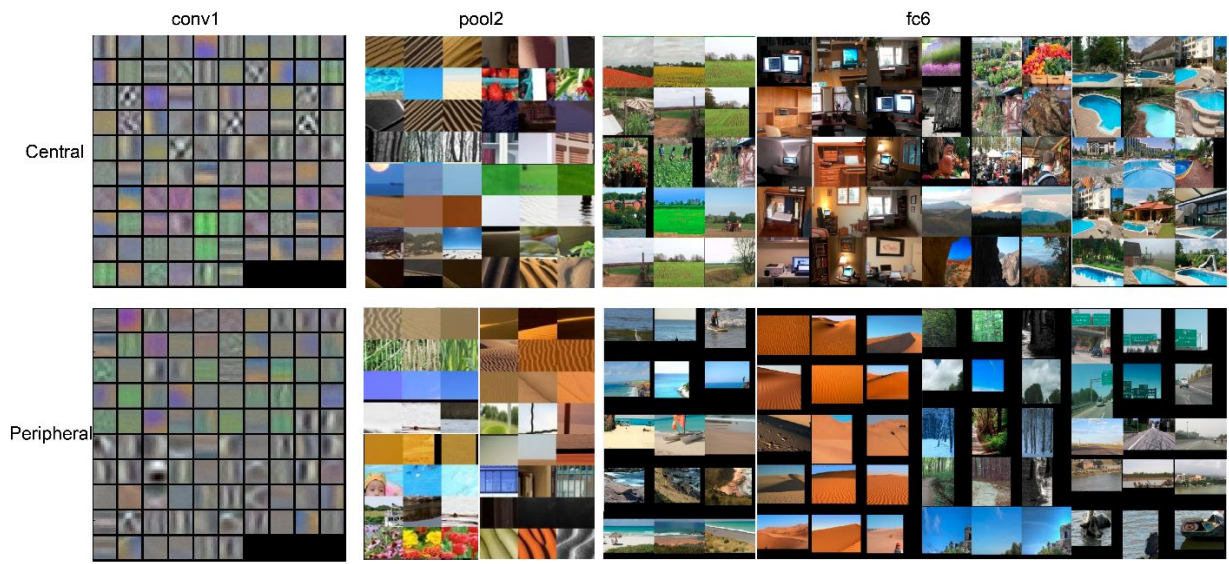

**Figure 9.** The left-hand column shows a visualization of the receptive fields of the first layer of features, while the remaining images display the three image patches that most highly activate various feature maps in *pool2* and individual units in *fc6* for the central (top row) and peripheral (bottom row) pathways of TDM.

From Figure 9, we can see the increasing complexity of features as the depth increases. For the *conv1* layer, both pathways learned the same V1 like features- edges of different spatial frequency and orientations, as well color opponency cells. For the *pool2* layer, both pathways learned more complicated features such as shapes or textures. For the *fc6* layer, the features respond to a much bigger receptive field size and even reveal categorical information to some extent. When comparing the features learned in the central pathway versus the peripheral pathway, it seems that they have different preferences over scene categories: central pathway prefers pools or indoor scenes, the peripheral pathway prefers desert and coast. Table 2 displays the exact number of features that belong to each scene category for the two pathways, selected by counting units for which the top 3 activating images are all from the same category.

| Category | Mean Central Weight | Central (C) | Peripheral (P) | Category | Mean Central Weight | C | P |
|---|---|---|---|---|---|---|---|
| **Pool** | 0.15 | **62.8** | 48.8 | **Highway** | 0.08 | 25.8 | **43.2** |
| **Market** | 0.13 | **14.5** | 7.2 | **River** | 0.07 | 4.8 | **22.4** |
| **Mountain** | 0.12 | **24.4** | 8.8 | **Forest** | 0.07 | 14.6 | **15.0** |
| **Indoor** | 0.09 | **33.4** | 12.2 | **Coast** | 0.06 | 6.0 | **27.0** |
| **Farm** | 0.09 | **27.0** | 11.0 | **Desert** | 0.05 | 8.2 | **33.0** |

**Table 2.** Summary of the average number of units in fc6 for which the top 3 images are all from the same category, along with the mean gating value for all images from that category. While the network always heavily weights the peripheral network, the two networks have allocated their representational resources differently.

Interesting findings can be seen from Table 2. For the central pathway, it learns more features that favor the following categories: pool, market, home, farm, and mountain, which are mostly man-made scene categories. The peripheral pathway favors forest, river, beach, desert, and street, which are mostly natural scene categories. The averaged gating node values also suggest that categories more preferred by the central pathway generally have higher mean weight than the categories more preferred by the peripheral pathway, although peripheral pathway still dominates the weight. This finding is consistent with our intuition, because man-made scene categories might contain lots of small objects and high-frequency details that need to be processed by central vision, and natural scene categories usually occupy the entire visual field and can be recognized using low spatial frequency contents and uniform global shapes that are usually processed by peripheral vision. These preferences of central and peripheral vision are also consistent with our hypothesis that central and peripheral vision generate different internal representations for scene recognition in TDM.

In summary, we showed the reason why peripheral vision is more important than central vision for scene recognition is due to two things: 1) peripheral vision contains better features for scene recognition than central vision, and 2) peripheral vision generates better internal representations (a spreading transform) than central vision, and that leads to better scene recognition performance.

## Discussion and Conclusions

Neurocomputational models are generally used to model and provide insight into behavioral data by proposing hypotheses about mechanisms that explain the data. One benefit for building these models is to analyze them in ways that are difficult or infeasible for humans, such as visualizing features and analyzing internal representations. Furthermore, we can put them in situations that are outside the normal biology, in order to understand the normal case better, such as using foveated images instead of log-polar ones, and training in conditions different from those experienced by humans.

In this work, we used deep CNN-based models to explore the contribution of central and peripheral vision for scene recognition. In particular, we modeled the behavioral result of Larson and Loschky (2009) and explained the data (Experiment 1). We trained deep CNNs on the same task as in the behavioral experiment, that is, to recognize ten different scene types, and then tested them under the Window and Scotoma conditions. We showed that, for all the deep CNN architectures we deployed, our results fit the human data very well: as the radius of visual angle increases, the recognition accuracy for the Window and Scotoma conditions increase and decrease monotonically. Importantly, we replicated the fact that the Scotoma condition achieves higher performance than the Window condition at or below $5°$, demonstrating that peripheral vision is better than central vision in maximizing scene recognition accuracy. Using log-polar transformed images to account for cortical magnification makes our results closer to the behavioral data. In addition, somewhat counterintuitively, our result coincided with the behavioral result that central vision is more efficient than peripheral vision on a per-pixel basis. Finally, we showed that our result is very robust, and not influenced by photographer bias.

In Experiment 2, we found that in our model, the predicted critical radius, where the central and peripheral pathways produce the same accuracy, is within the measured tolerance of the human experiments. The model's critical radius is 8.00°, which is within the 95% confidence interval of the measured critical radius of 7.48°. All of these results demonstrated that our deep CNN-based models are plausible models to simulate and explain the findings related to scene perception in humans.

We then used our models to explain *why* peripheral vision contributes more to scene recognition than central vision, and to predict *how* it is achieved in cortex. We proposed two hypotheses: 1) The features contained in peripheral vision are better for scene recognition than central vision. 2) The internal representation that peripheral vision generates is better than central vision for scene recognition. We designed two experiments to test the two hypotheses. In the first experiment (Experiment 3.1), we used single pathway deep CNNs that are trained only on Window or Scotoma images to test whether the loss of peripheral vision or central vision is vital for scene recognition. This experiment is complementary to Experiments 1 and 2, where we used full images to train the network, and Window and Scotoma images to test the network. Here, we used Window/Scotoma images to train the network and full images to test the network. The question we are trying to answer in Experiment 3.1 is: if we were born without central or peripheral vision, what will happen to scene recognition performance? If peripheral vision contains better feature than central vision for recognizing scenes, then learning without central vision (i.e., the Scotoma condition) should not impair the recognition as severely as learning without peripheral vision (i.e,, the Window condition). Our experimental result showed that having peripheral vision produced similar performance as having the full range of vision, but having only central vision causes significant loss in recognition performance. This result again demonstrated the superior usefulness of peripheral vision.

For the second hypothesis, that the internal representation of peripheral vision is better than that used by central vision for scene recognition (Experiment 3.2), we used a mixture-of-experts version of TDM to build a developmental model of scene processing pathways in the human visual system. We ran different experiments to analyze the internal representations in TDM. By analyzing the value of the gating nodes in TDM, we showed that TDM heavily weights the peripheral pathway over the central pathway when trying to categorize scenes, suggesting the superior representation that peripheral vision generated during the scene recognition process. By plotting the gating node value as a function of training iterations, we can see a clear increasing trend for the node corresponding to the peripheral pathway, which is consistent with the trend of improving scene recognition accuracy. Mapping this process into the human developmental process, we can hypothesize that even if the central and peripheral pathways started equal, the consistent advantage of peripheral information gradually shapes the network to lean towards the peripheral pathway to recognize scenes. The peripheral advantage emerges naturally during the development of the scene recognition system, and remains stable throughout its maturity. We also showed that when the weights between the two pathways are fixed to be equal, the network learns to turn off central vision.

We also visualized the internal representation of the *fc6* layer of the two pathways by projecting its net input across all validation images into a 2-D subspace using PCA. We found that peripheral pathway produces a more distinct clustering of the different categories than the central pathway, which appears to clump all of the categories together. Further visualization of the *fc6* features suggest that the two pathways have different preferences over scene categories. Remarkably, even over several runs with different initial random weights, there is a consistent mapping of preferences for different scene categories to each pathway. This remains to be explained and demands further replication.

We designed our model as an instantiation of the anatomical separation between the central and peripheral pathways - one leading from foveal input to the Lateral Occipital Complex (LOC), and the other from peripheral input to the Parahippocampal Place Area (PPA). It is well known that topographical cortical representations are revealed in the retinotopic visual areas, where mapping the eccentricity and phase angle components of the retinotopic map results in iso-eccentricity bands orthogonal to the meridian representations of the angles. In higher order object-related visual areas, multiple studies (Hasson et al., 2003; Levy et al., 2001; Malach et al., 2002; Grill-Spector & Malach, 2004) have shown that orderly central and peripheral representations can still be found in regions engaged in face and place perception. In particular, the FFA is enervated by foveal vision, and the PPA is enervated by peripheral vision (Arcaro et al., 2009; Nasr et al., 2011; Grill-Spector & Weiner, 2014). These studies further hypothesize that the cortical topography provides a global organizing principle of the entire visual cortex. Recent anatomical studies on white-matter connectivity and cytoarchitecture, as well as functional neuroimaging studies for object areas in the ventral temporal cortex have shown that the central and peripheral representations are segregated by the mid-fusiform sulcus (Weiner et al., 2014; Grill-Spector & Weiner, 2014; Gomez et al., 2015; Lorenz et al., 2015). Following this organizing principle of parallel processing of central and peripheral visual information, TDM is its direct application to providing a more realistic model for the scene recognition system.

The peripheral advantage for scene recognition in our model supports the importance of PPA in scene recognition. The PPA is activated more for buildings and scenes than other categories, such as faces (Epstein et al., 1999), and it is involved in scene memory (Ranganath, DeGutis, & D'Esposito, 2004; Brewer, Zhao, Desmond, Glover, & Gabrieli, 1998). The PPA has also been shown to respond to the spatial layout or geometry of the scene -- that is, whether a scene is "open" versus "closed" (Oliva & Torralba, 2001), in tasks like navigation (Epstein & Kanwisher, 1998; Janzen & Van

Turennout, 2004) and scene classification (Park, Brady, Greene, & Oliva, 2011). However, PPA activity is not modulated by the number of objects in the scene (Epstein & Kanwisher, 1998). Many studies using multivoxel pattern analysis (MVPA) have shown that the fMRI activity of PPA can be decoded and used to distinguish between different scene categories (Naselaris, Prenger, Kay, Oliver, & Gallant, 2009; Walther, Caddigan, Fei-Fei, & Beck, 2009; Park et al., 2011). One thing to note is that although we showed peripheral vision contributes more to scene recognition than central vision, we did not ignore the fact that central vision (or LOC according to our mapping) may actually have a role in scene recognition. In our visualization experiment (Figure 9), we showed that the central pathway has more features characteristic of man-made scene categories, and this may be because the particular objects in those categories (such as a desk in a home, and vegetables and fruit in the market category) are important to distinguish them from other categories. In fact, since LOC is specialized in representing object shapes and object categories (Grill-Spector, Kushnir, Edelman, Itzchak, & Malach, 1998), it stands to reason that LOC should be encoding the content of a scene when there are objects presented in the scene. In fact, the pattern of neural responses in the LOC has also been shown to differentiate among scene categories (Walther et al., 2009; Park et al., 2011) and decode whether certain objects were presented within the scenes (Peelen, Fei-Fei, & Kastner, 2009).

What is the peripheral information that contributes to scene recognition? Since the periphery mainly contains low-spatial frequency information, it is natural to argue that low resolution coarse information is the key. In fact, it is well-known that scene perception follows a coarse-to-fine processing paradigm (Schyns, P. G., & Oliva, A., 1994), that is, low-spatial frequency (LSF) information dominates scene categorization when the presentation is very short (30 ms), but high spatial frequency (HSF) information dominates later (150 ms). Other behavioral studies also suggest LSF-based processing during rapid scene recognition (Kihara & Takeda, 2010; De Cesarei & Loftus, 2011), and this preference emerges in the very early stages of development in 7-to-8-month-old infants (Otsuka, Ichikawa, Kanazawa, Yamaguchi, & Spehar, 2014). However, more careful manipulation of spatial frequencies and time-course analysis is needed to elucidate the interaction between spatial frequency processing and scene recognition performance. Is the dominance of high frequency information later due to input from the LOC in the scene categorization process?

As a neurocomputational model, TDM is generic and can be applied to a broader range of modeling tasks beyond the present study. For scene perception, it is possible to incorporate recent findings into TDM, such as the role of color and modified images in peripheral vision (Nuthmann & Malcolm, 2016; Wallis, Bethge, & Wichmann, 2016). TDM can also be integrated with other retinal or encoding models (Ehinger & Rosenholtz, 2016, Chessa, Maiello, Bex, & Solari, 2016, Shan, Tong, & Cottrell, 2016) to build a more realistic model for human scene perception. Following the organizing principle of central-peripheral representation across the visual cortex, it is natural to incorporate other important object categories that are associated with central vision in the VTC, such as objects (LOC), faces (FFA), and words (VWFA), into our model to explore their interactions with scene recognition, and to test whether central and peripheral preferences for these categories can be found in TDM. In addition, it is possible to extend TDM to model other organizing principles of the brain, for example, the left-right hemispheric asymmetry. It is well known that right hemisphere (RH) is lateralized for the processing of LSF global information, and left hemisphere (LH) is lateralized for the processing of HSF local information (Sergent, 1982). For scene perception, it has been shown that LSF scenes are recognized faster in RH than LH, and HSF scenes are recognized faster in LH than RH (Peyrin, Chauvin, Chokron, & Marendaz, 2003). In a neuroimaging study, Peyrin, Baciu, Segebarth, & Marendaz (2004) showed that scene perception was based mainly on LSF analysis in the right hemisphere by showing significant activations in right PPA. Musel et al. (2013) further investigated the interaction of retinotopy and the functional lateralization of spatial frequency processing of scene categorization, and provided the first evidence of retinotopic processing of spatial frequencies: LSF information elicited activation associated with peripheral visual field, and HSF information elicited activation associated with the fovea. This retinotopic spatial frequency processing, as well as hemispheric lateralized processing of scenes may provide a unified theory for scene and object recognition in visual cortex.

More recently, another organizing principle of the visual cortex, that is, the upper (ventral surface) and lower (lateral surface) visual field organization, has been proposed (Silson, Chan, Reynolds, Kravitz, & Baker, 2015). Using fMRI studies, Silson et al. (2015) showed that a strong bias of population receptive field mapping for the contralateral upper and lower quadrant can be found within the ventral (PPA) and lateral (transverse occipital sulcus, or TOS, Dilks, Julian, Paunov, & Kanwisher, 2013) scene-selective regions, respectively. Extending our model to incorporate these constraints would enable testing potential biases in different quadrants of the visual field that might result from these anatomical constraints.

In summary, we suggest that the advantage of peripheral vision over central vision in scene recognition is due to the intrinsic usefulness of the features carried by peripheral vision, and it helps to generate a greater spreading transform in the internal representational space that enables better processing for scene categories. The peripheral advantage emerges naturally as a developmental process of the visual system. Furthermore, we predict that the two pathways correlate with their neural substrates of LOC and PPA, and both contribute to an integrated scene recognition system.

## Acknowledgements

We thank the members of the Perceptual Expertise Network (PEN) and Gary's Unbelievable Research Unit (GURU) for comments on this work. PW was supported in part by a gift from Hewlett-Packard. GWC was supported in part by NSF grant